\documentclass[aip,reprint]{revtex4-1}
\usepackage[utf8]{inputenc}
\usepackage[T1]{fontenc}
\usepackage{textcomp}
\usepackage{threeparttable}
\usepackage{hyperref}
\usepackage{graphicx}
\setcitestyle{numbers,super,sort&compress}

\begin{document}

\title{Building an Affordable Self-Driving Lab: Practical Machine Learning Experiments for Physics Education Using Internet-of-Things}

\author{Yang Liu}%
\thanks{These authors contributed equally to this work.}
\affiliation{School of Advanced Materials and Nanotechnology,
             Xidian University, Xi’an 710071, China}

\author{Qianjie Lei}%
\thanks{These authors contributed equally to this work.}
\affiliation{School of Advanced Materials and Nanotechnology,
             Xidian University, Xi’an 710071, China}
\author{Xiaolong He}
\thanks{These authors contributed equally to this work.}
\affiliation{School of Advanced Materials and Nanotechnology,
             Xidian University, Xi’an 710071, China}
\author{Yizhe Xue}
\author{Kexin He}
\author{Haitao Yang}
\author{Yong Wang}
\author{Xian Zhang}
\author{Li Yang}
\author{Yichun Zhou}
\email{yichunzhou@xidian.edu.cn}
\affiliation{School of Advanced Materials and Nanotechnology, Xidian University, Xi’an 710071, China}

\author{Ruiqi Hu}
\affiliation{Department of Materials Science and Engineering, University of Delaware, Newark, Delaware 19716, United States}

\author{Yong Xie}
\email{yxie@xidian.edu.cn}
\affiliation{School of Advanced Materials and Nanotechnology, Xidian University, Xi’an 710071, China}
\affiliation{2D Foundry, Instituto de Ciencia de Materiales de Madrid (ICMM-CSIC), Madrid, E-28049, Spain}

\date{\today}

\begin{abstract}

Machine learning (ML) is transforming modern physics research, but practical, hands-on experience with ML techniques remains limited due to cost and complexity barriers. To address this gap, we introduce an affordable, autonomous, Internet-of-Things (IoT)-enabled experimental platform designed specifically for applied physics education. Utilizing an Arduino microcontroller, a customizable multi-wavelength light emitting diode (LED) array, and photosensors, our setup generates diverse, real-time optical datasets ideal for training and evaluating foundational ML algorithms, including traversal methods, Bayesian inference, and deep learning. The platform facilitates a closed-loop, self-driving experimental workflow, encompassing automated data collection, preprocessing, model training, and validation. Through systematic performance comparisons, we demonstrate the superior ability of deep learning to capture complex nonlinear relationships compared to traversal and Bayesian methods. At approximately ~\$60, this open-source IoT platform provides an accessible, practical pathway for students to master advanced ML concepts, promoting deeper conceptual insights and essential technical skills required for the next generation of physicists and engineers.

\end{abstract}

\maketitle

%=====================================================================
\section{Introduction}
\label{sec:intro}

Artificial intelligence (AI) has revolutionized modern research across physics and engineering disciplines~\cite{nobel2024Phys,nobel2024Chem, Bennett2024,Tom2024}. Recent advances demonstrate AI’s pivotal role in accelerating the synthesis of novel materials~\cite{Bennett2024,Tom2024,Ament2021,Shimizu2020}, their characterization~\cite{Liu2022,Slautin2024,Masubuchi2018,Yang2024}, and the emergence of autonomous or "self-driving" experimental laboratories~\cite{YunchaoXie2023,Burger2020,Szymanski2023,Adam2024,DelgadoLicona2025,Waelder2025}. For example, machine learning (ML) and deep learning (DL) approaches are actively deployed to facilitate rapid discovery of next-generation battery materials and elucidate intricate protein structures~\cite{nobel2024Chem}. In material characterization, integrating DL techniques with computer vision has significantly improved the precision in determining layer thickness~\cite{Masubuchi2018} and twist angles in two-dimensional materials~\cite{Yang2024}.  
Beyond characterization, the integration of robotics with AI has further advanced the synthesis and transfer of two-dimensional materials, as demonstrated in recent works on automated growth control and material handling~\cite{Booth2025,Zhao2025,Waelder2025}.
 Furthermore, by coupling density-functional theory (DFT) and ML algorithms, researchers can efficiently predict novel crystal structures and their electronic properties, thereby streamlining materials innovation~\cite{Hou2024,Xiao2023,Horton2025}. The recent Nobel Prize in Physics, awarded to John J. Hopfield and Geoffrey E. Hinton, further underscores the transformative impact of neural network methodologies in scientific research~\cite{nobel2024Phys}.

Parallel to AI advancements, the Internet of Things (IoT) has emerged as an essential technology for real-time data acquisition, processing, and integration across various sectors, driven by its affordability, adaptability, and compatibility with widely used programming environments such as Python and MATLAB. Affordable IoT frameworks, paired with inexpensive sensor networks and new semiconducting sensing materials, have dramatically expanded experimental capabilities in fields from healthcare diagnostics to environmental monitoring~\cite{Zhang2023_OEA,Xie17,Xie2020,Zhang2023_TrendsChem,Sozen2023, Xie2012_JAP}. In STEM education, cost-effective and open-source experimental tools such as LEGO-based autonomous platforms are democratizing access to advanced laboratory experiences~\cite{Baird2022,Saar,ginsburg2023}. The proliferation of interconnected sensor technologies combined with AI analytics is now transforming traditional research environments into adaptive, self-driving laboratories capable of autonomous decision-making and iterative learning~\cite{Zanella2014}.

In this paper, we present a novel and cost-effective IoT-based experimental platform designed explicitly for applied physics education. This platform integrates foundational ML algorithms—traversal, Bayesian inference, and deep learning—into practical, hands-on exercises. By employing a modular setup utilizing Arduino microcontrollers, customizable light emitting diode (LED) arrays, and multispectral sensor arrays, students engage directly with complete ML workflows, from real-time data acquisition and preprocessing to model training, prediction, and validation. The practical experiments conducted in this self-driving laboratory environment provide clear demonstrations of each algorithm's effectiveness, strengths, and limitations, significantly enhancing students' conceptual understanding and technical proficiency.

This accessible yet powerful approach equips emerging physicists and engineers with critical problem-solving skills and practical expertise, highlighting the transformative potential of combining AI and IoT technologies to address complex scientific challenges.

%--------------------------------------------------------------------

\section{Experimental Setup}
\label{sec:setup}
\begin{figure*}[t]
  \centering
  \includegraphics[width=\textwidth]{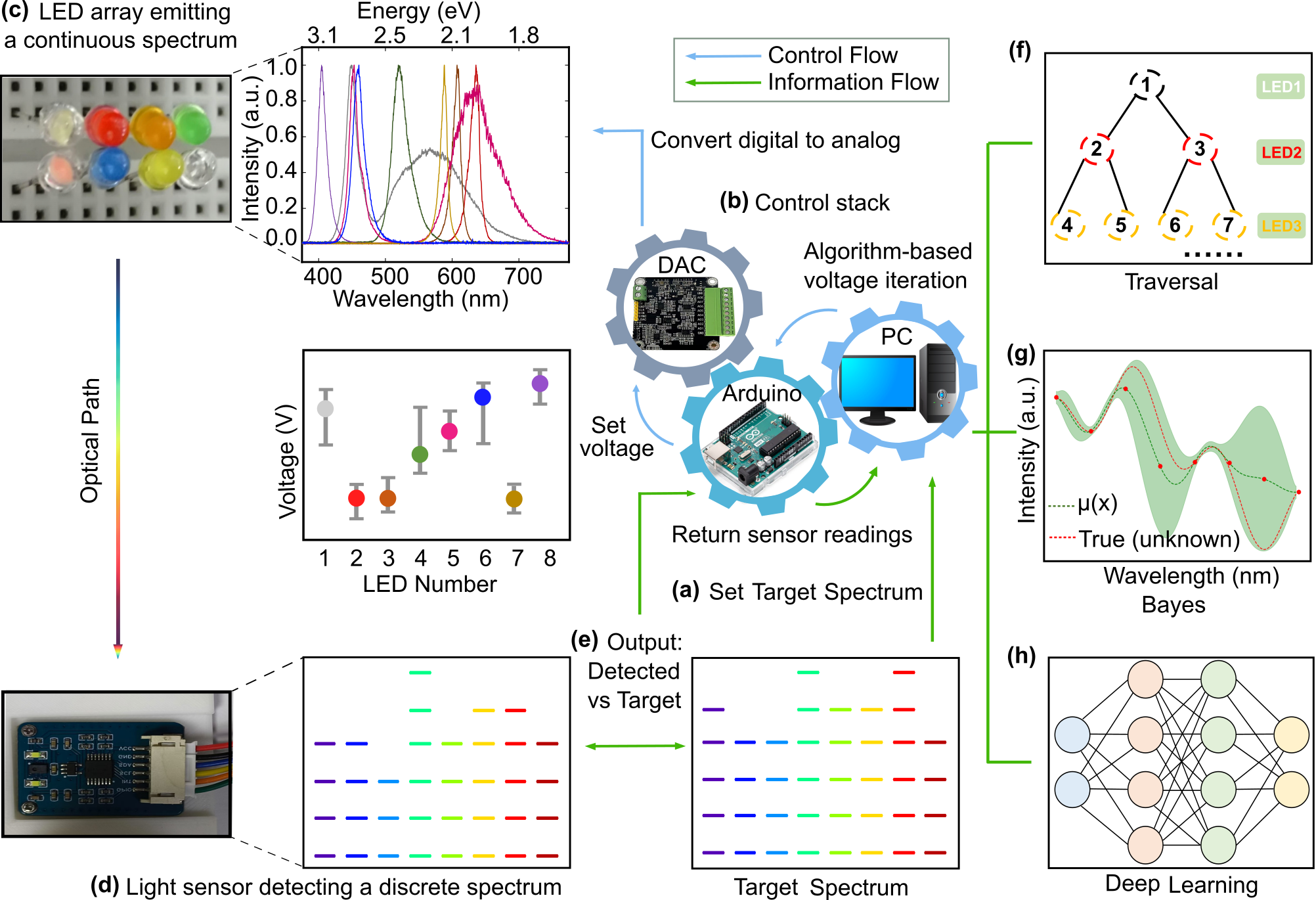}
  \caption{Schematic of the affordable Internet-of-Things (IoT) platform for closed-loop, self-driving optical experiments.  
  \textbf{(a)}~The user defines a target spectrum to be reproduced.  
  \textbf{(b)}~Control stack consisting of PC, Arduino, and DAC for algorithm-based voltage iteration and digital-to-analog conversion.  
  \textbf{(c)}~LED array emitting a continuous spectrum under voltage modulation.  
  \textbf{(d)}~Multichannel light sensor (AS7341) detecting the optical output as a discrete spectrum.  
  \textbf{(e)}~Comparison between detected and target spectra closes the feedback loop.  
  The right column illustrates the three algorithmic controllers used:  
  \textbf{(f)}~Traversal search through a hierarchical tree of candidate voltages.  
  \textbf{(g)}~Bayesian optimization using a probabilistic surrogate model $\mu(x)$ with confidence intervals.  
  \textbf{(h)}~Deep learning controller implemented via a neural network for voltage prediction.  
  Blue arrows represent control flow (voltage commands), while green arrows represent information flow (measured spectra).}
  \label{fig:1}
\end{figure*}

The experimental setup of the self-driving IoT platform was designed to be affordable, modular, and accessible, relying exclusively on commercially available components and open-source software.  
Figure~\ref{fig:1} illustrates the overall architecture of the system.

In Fig.~\ref{fig:1}(c), an array of eight light-emitting diodes (LEDs) with distinct emission wavelengths serves as the tunable optical source for dataset generation.  
By adjusting the driving voltages, the intensity and spectral distribution of each LED can be precisely modulated, enabling construction of diverse optical outputs required for training and validation.  

The control stack is shown in Fig.~\ref{fig:1}(b).  
An Arduino microcontroller provides real-time hardware interfacing, handling communication with the peripheral devices and ensuring robust execution of control signals.  
A multi-channel digital-to-analog converter (DAC) translates the digital voltage commands into analog signals to drive the LEDs.  
The host computer executes the optimization algorithms, compares detected and target spectra, and iteratively updates the voltage setpoints.  
When required, GPU or cloud resources can be leveraged to accelerate algorithm execution and data processing.

Spectral feedback is acquired by a multispectral light sensor (AS7341, Adafruit) [Figs.~\ref{fig:1}(d) and \ref{fig:3}(e)]~\cite{AdafruitAS7341}.  
The device provides ten independent detection channels: eight covering narrow visible bands from approximately 415~nm to 680~nm, complemented by broadband “clear” and near-infrared channels.  
An integrated analog-to-digital converter and I$^2$C interface facilitate compact, high-resolution measurements, making the sensor particularly suitable for spectral characterization in low-cost educational or research platforms.

On the algorithmic side [Figs.~\ref{fig:1}(f)–\ref{fig:1}(h)], three control strategies were implemented.  
The traversal algorithm systematically explores the voltage parameter space to minimize spectral mismatch.  
The Bayesian algorithm incorporates prior knowledge and uncertainty quantification through a probabilistic surrogate model.  
Finally, the deep-learning controller employs a convolutional neural network trained on synthetic and experimental datasets to directly predict optimal voltages from target spectra.

% \begin{figure*}[t]
%   \centering
%   \includegraphics[width=\textwidth]{Figures/Fig2.png}
%   \caption{Components and configurations of the IoT setup.
%   (a) Hardware list.  
%   (b) Sensor‑array wiring.  
%   (c) LED‑sensor alignment.}
%   \label{fig:2}
% \end{figure*}

% Figure~\ref{fig:2} lists all components: (1) sensor array; (2) DAC board; (3) Arduino; (4) 3D‑printed holders; (5) LEDs; (6) jumper wires; (7) breadboard; (8) USB cable.  

\begin{figure*}[t]
  \centering
  \includegraphics[width=\textwidth]{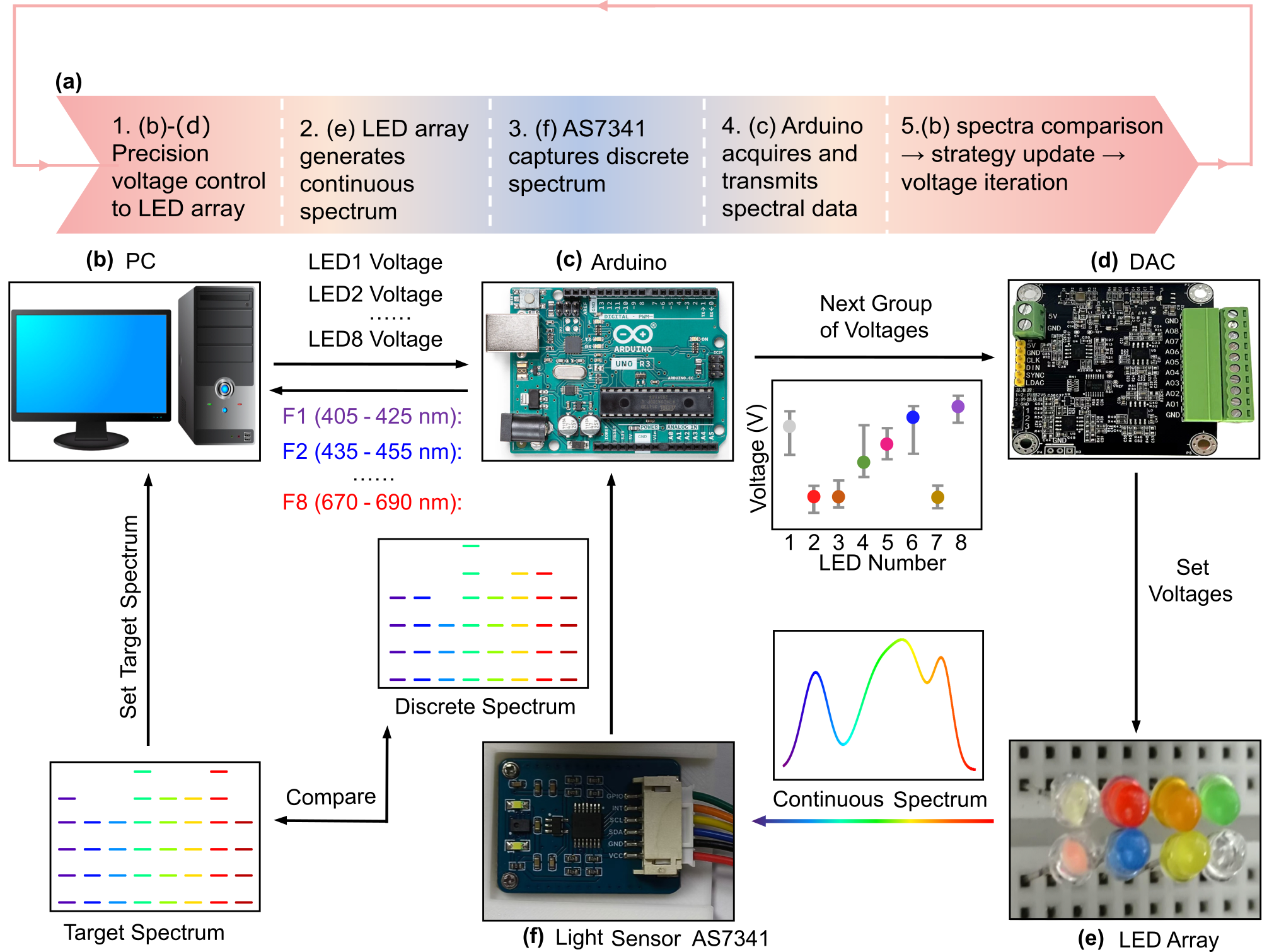}
  \caption{System workflow of the closed-loop IoT-based platform for self-driving optical experiments.  
  \textbf{(a)}~Overall workflow: precision voltage control applied to the LED array, generation of continuous spectra, multispectral sensing of discrete bands, Arduino-based data acquisition and transmission, and iterative comparison with the target spectrum for algorithm-guided updates.  
  \textbf{(b)}~Host computer (PC) defining the target spectrum and executing optimisation algorithms.  
  \textbf{(c)}~Arduino microcontroller serving as the central hardware interface for communication and control.  
  \textbf{(d)}~Multi-channel DAC converting digital commands into analog voltages for the LED array.  
  \textbf{(e)}~LED array emitting broadband spectra under voltage modulation.  
  \textbf{(f)}~AS7341 multispectral light sensor measuring the discrete spectral response (F1–F8) and returning data to the host for comparison with the predefined target.}
  \label{fig:2}
\end{figure*}

Figure~\ref{fig:2} illustrates the closed-loop workflow of the platform.  
The host computer defines the target spectrum, executes the optimization algorithm, and generates candidate voltage vectors [Fig.~\ref{fig:2}(a)].  
Through the Arduino microcontroller [Fig.~\ref{fig:2}(c)], these commands are converted by the DAC into analog voltages [Fig.~\ref{fig:2}(d)] to drive the LED array [Fig.~\ref{fig:2}(e)].  
The emitted light is measured by the AS7341 multispectral sensor [Fig.~\ref{fig:2}(f)], which returns discrete spectral components for comparison against the predefined target.  
This feedback closes the control loop, enabling algorithm-guided iterative refinement of the LED outputs.  

The hardware was assembled using standard prototyping practices, with firmware developed in the Arduino IDE and higher-level algorithms implemented in Python leveraging \texttt{NumPy} and \texttt{PyTorch}.  
Such a modular hardware–software framework offers a flexible testbed for machine-learning-driven experimentation and can be seamlessly extended toward fully autonomous laboratory operation through large-language-model (LLM)–assisted instrumentation programming~\cite{Xie2025_sstr}, as recently demonstrated in single-pixel imaging and scanning photocurrent microscopy.\\
The hardware components and their key specifications are summarized in Table~\ref{tab:hardware}.

\begin{table*}[htbp]
\caption{\label{tab:hardware} Hardware components and key specifications.}
\begin{threeparttable}
\begin{ruledtabular}
\begin{tabular}{p{1.5cm} p{1.5cm} p{2.5cm} p{5cm} p{1.7cm} p{1cm}}
\textbf{Device} & \textbf{Model} & \textbf{Key specs} & \textbf{Role} & \textbf{Supplier} & \textbf{Cost (\$)} \\
\hline
Light sensor & AS7341 & I$^{2}$C interface, 3.3V/5V & 8-channel visible light sensor & Waveshare\tnote{1} & 15.32 \\
DAC & AD5328 & 8-channel output, SPI, 5V & Precision voltage output for LED control & Keyi\tnote{2} & 14.48 \\
Arduino & Uno Rev3 & Operating voltage: 5V & Microcontroller for data acquisition and communication. Receives voltage commands and transmits sensor readings & Arduino\tnote{3} & 34.28 \\
3D-printed holders & Self-developed & Figure S2 & Secure sensor above LED for convenient data acquisition & Self-developed & -- \\
LED & LED chip & Width: 3\,mm & Programmable visible light source for spectral response testing & Zave\tnote{4} & 2.09 \\
Jumper wires & Male-to-Female & Pitch: 2.54\,mm, Rated 300V & Modular interconnection with pluggable terminals & Risym\tnote{5} & 1.31 \\
Breadboard & 830-hole & Dimensions: 165$\times$55\,mm (±0.5\,mm) & Expandable solderless breadboard & Risym\tnote{6} & 0.74 \\
USB cable & US135 & USB Type-B male plug & Connects Arduino Uno to PC for bidirectional serial communication & UGREEN\tnote{7} & 1.39 \\
\end{tabular}
\end{ruledtabular}

\begin{tablenotes}
\item [1] \url{https://detail.tmall.com/item.htm?id=640012536606}
\item [2] \url{https://item.taobao.com/item.htm?id=752440272451}
\item [3] \url{https://store.arduino.cc/collections/uno/products/arduino-uno-rev3}
\item [4] \url{https://detail.tmall.com/item.htm?id=610942801333}
\item [5] \url{https://detail.tmall.com/item.htm?id=14466195609}
\item [6] \url{https://detail.tmall.com/item.htm?id=16513870165}
\item [7] \url{https://detail.tmall.com/item.htm?id=562303388266}
\end{tablenotes}
\end{threeparttable}
\end{table*}

%---------------------------------------------------------------------
\subsection{Data Collection and Pre‑processing}
\label{subsec:data}

Sensor readings were recorded while systematically varying LED voltages.  
Traversal and Bayesian experiments calibrated the sensitivity matrix $S$, which maps LED voltages to sensor outputs.  
For DL, a large synthetic dataset (100 000 samples) was generated, then partitioned into training, validation and test splits.  
All data were normalized, and outliers were removed prior to model training.

%---------------------------------------------------------------------
\subsection{Algorithm Implementation}
\label{subsec:algo}

Traversal uses iterative error minimization; Bayesian inference refines a probabilistic voltage distribution; DL employs a regression‑style CNN trained with Adam optimization.  
Algorithm accuracy was quantified via mean‑squared error (MSE) between inferred and target voltages.

%---------------------------------------------------------------------
\subsection{Experimental Procedure}
\label{subsec:procedure}

  \begin{figure*}[t]
  \centering
  \includegraphics[width=0.9\textwidth]{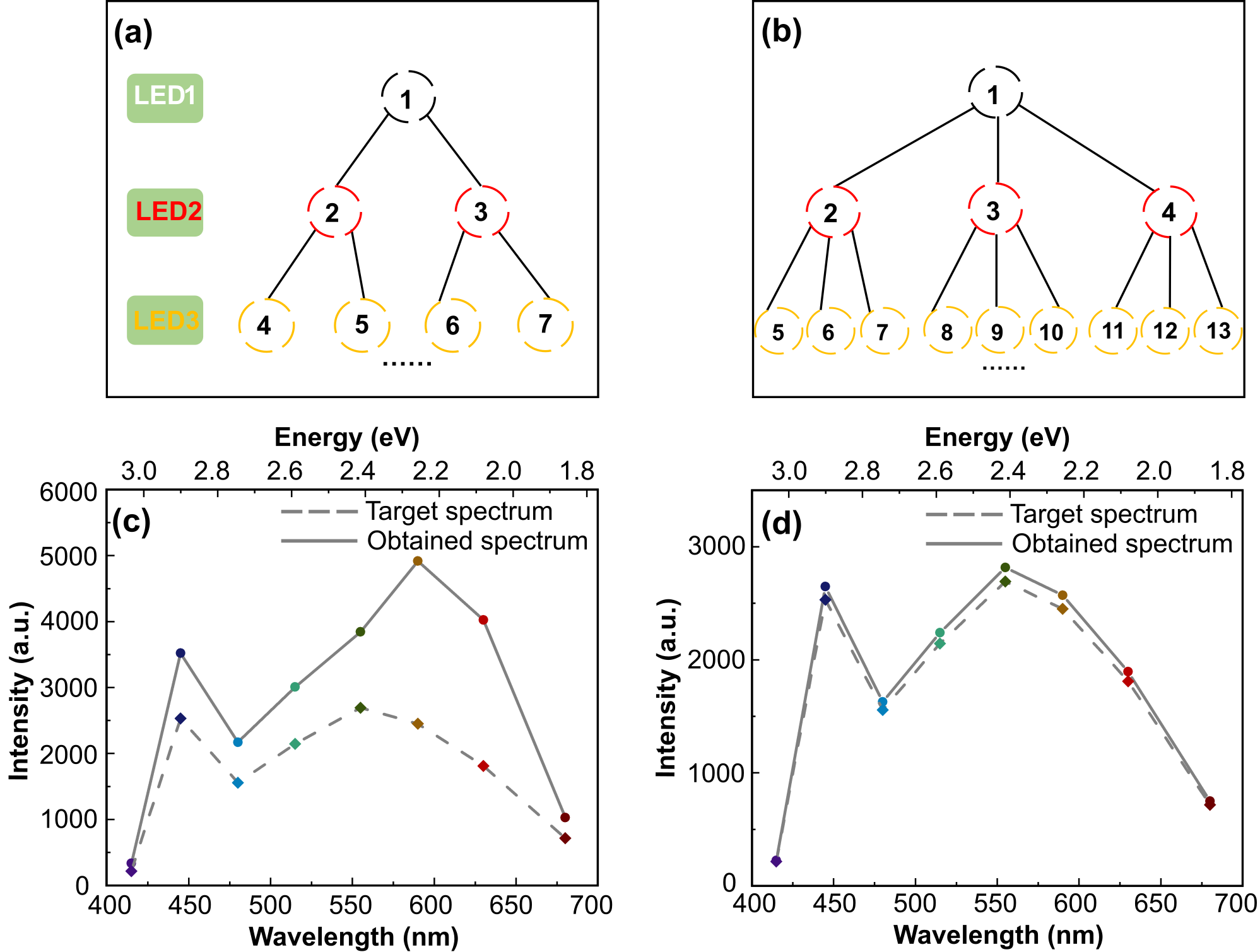}
  \caption{Traversal algorithm for voltage-space exploration and spectral matching.  
  \textbf{(a)}~Hierarchical tree representation of the traversal strategy, where each node corresponds to a candidate voltage vector and branches denote incremental updates across individual LED channels.  
  \textbf{(b)}~An expanded search tree with increased branching depth provides finer sampling of the multidimensional voltage space, mitigating the risk of trapping in local optima.  
  \textbf{(c)}~Experimental spectrum obtained with a coarse voltage step of 0.1\,V, showing only partial alignment with the target spectrum.  
  \textbf{(d)}~Spectrum obtained with a finer step size of 0.05\,V, yielding substantially improved agreement with the target and highlighting the benefit of enhanced search resolution.}
  \label{fig:3}
\end{figure*}

\begin{figure*}[t]
  \centering
  \includegraphics[width=\textwidth]{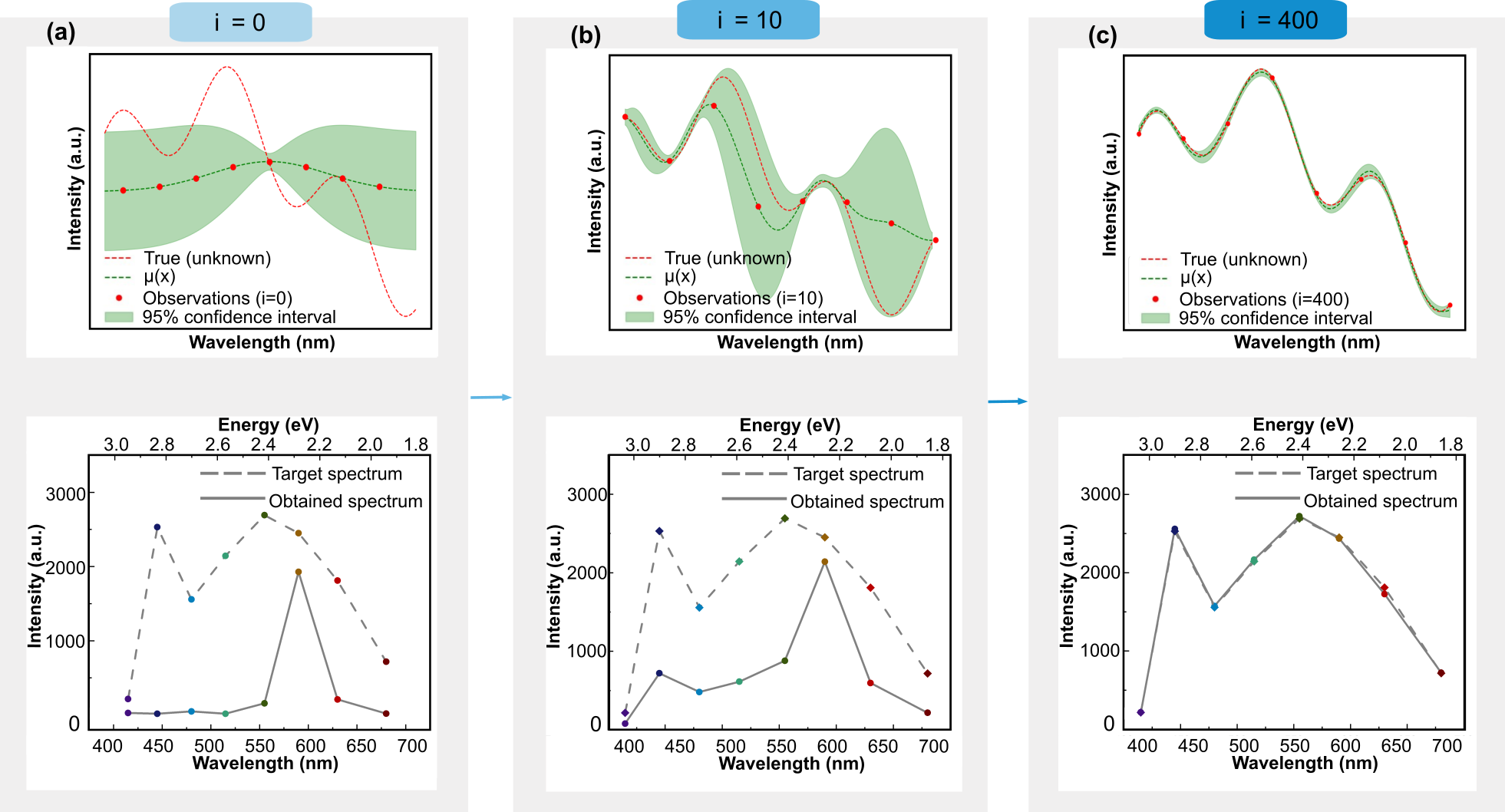}
  \caption{Bayesian optimization for closed-loop spectral control.  
  \textbf{(a)}~Initial state ($i=0$): the surrogate Gaussian process model $\mu(x)$, trained with a single observation, yields only a rough estimate of the spectrum–voltage relation and exhibits large predictive uncertainty (shaded 95\% confidence interval).  
  \textbf{(b)}~After 10 iterations ($i=10$), the model is updated with additional voltage–spectrum samples, reducing uncertainty and improving agreement with the true spectral response.  
  \textbf{(c)}~After 400 iterations ($i=400$), the surrogate converges closely to the underlying system response. Bottom panels show experimental spectra (solid lines) compared with the target spectra (dashed lines) at each stage, highlighting progressive convergence toward the desired output.}
  \label{fig:4}
\end{figure*}

\begin{figure*}[t]
  \centering
  \includegraphics[width=\textwidth]{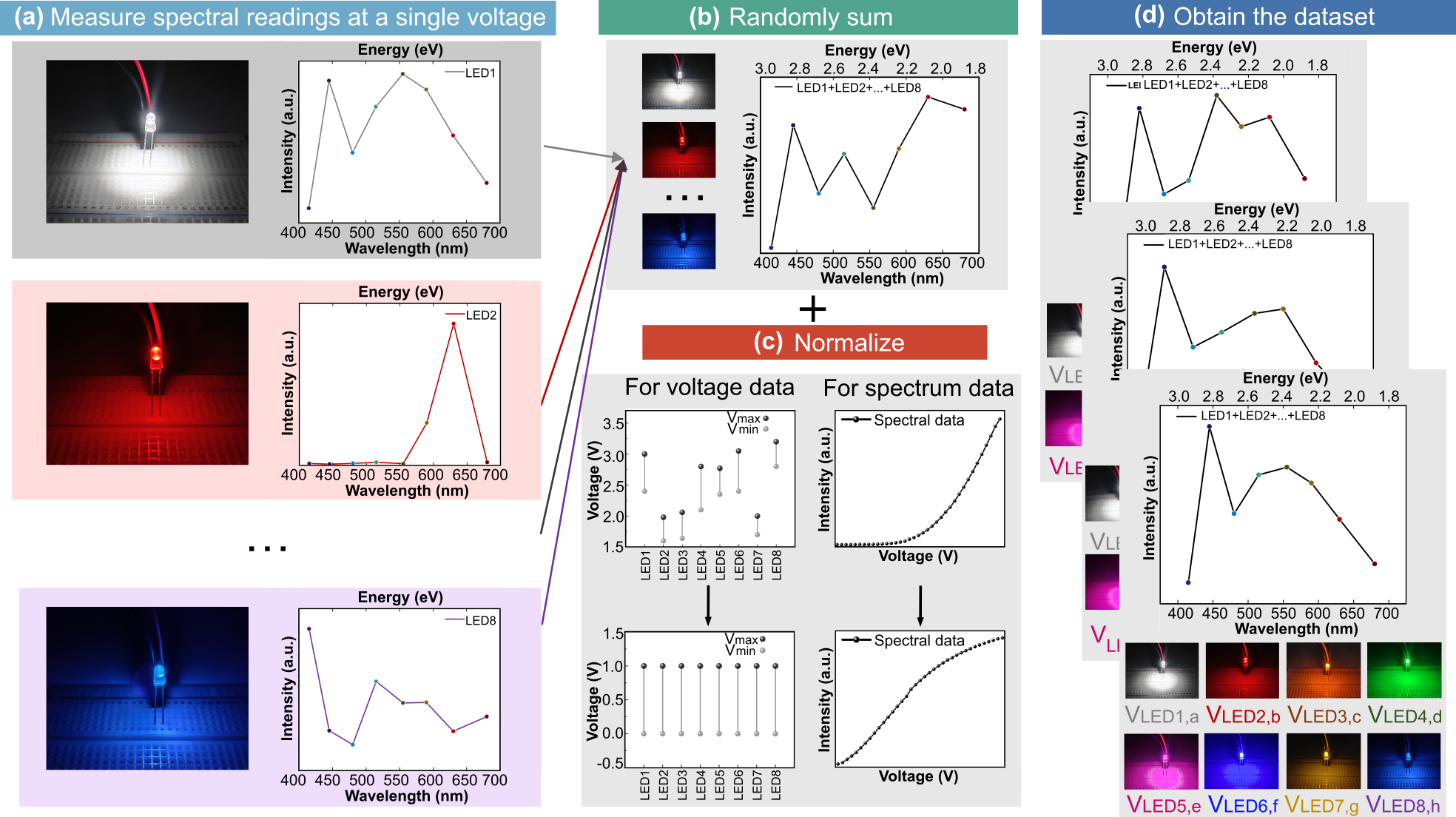}
  \caption{Dataset construction and preprocessing pipeline for training the deep learning controller.  
  \textbf{(a)}~Spectral responses of individual LEDs measured at discrete voltage levels. Each LED is driven independently, and its emission spectrum is recorded by the multispectral sensor.  
  \textbf{(b)}~Synthetic composite spectra are generated by randomly summing responses from different LEDs corresponding to randomly sampled voltage vectors.  
  \textbf{(c)}~Normalization of both input voltages and spectral outputs ensures consistent scaling and accelerates convergence during network training.  
  \textbf{(d)}~Final dataset consisting of 100,000 voltage–spectrum pairs, each encoding an 8-dimensional voltage vector and the corresponding composite spectrum, ready for supervised deep learning.}
  \label{fig:5}
\end{figure*}

\begin{figure*}[htbp]
  \centering
  \includegraphics[width=\textwidth]{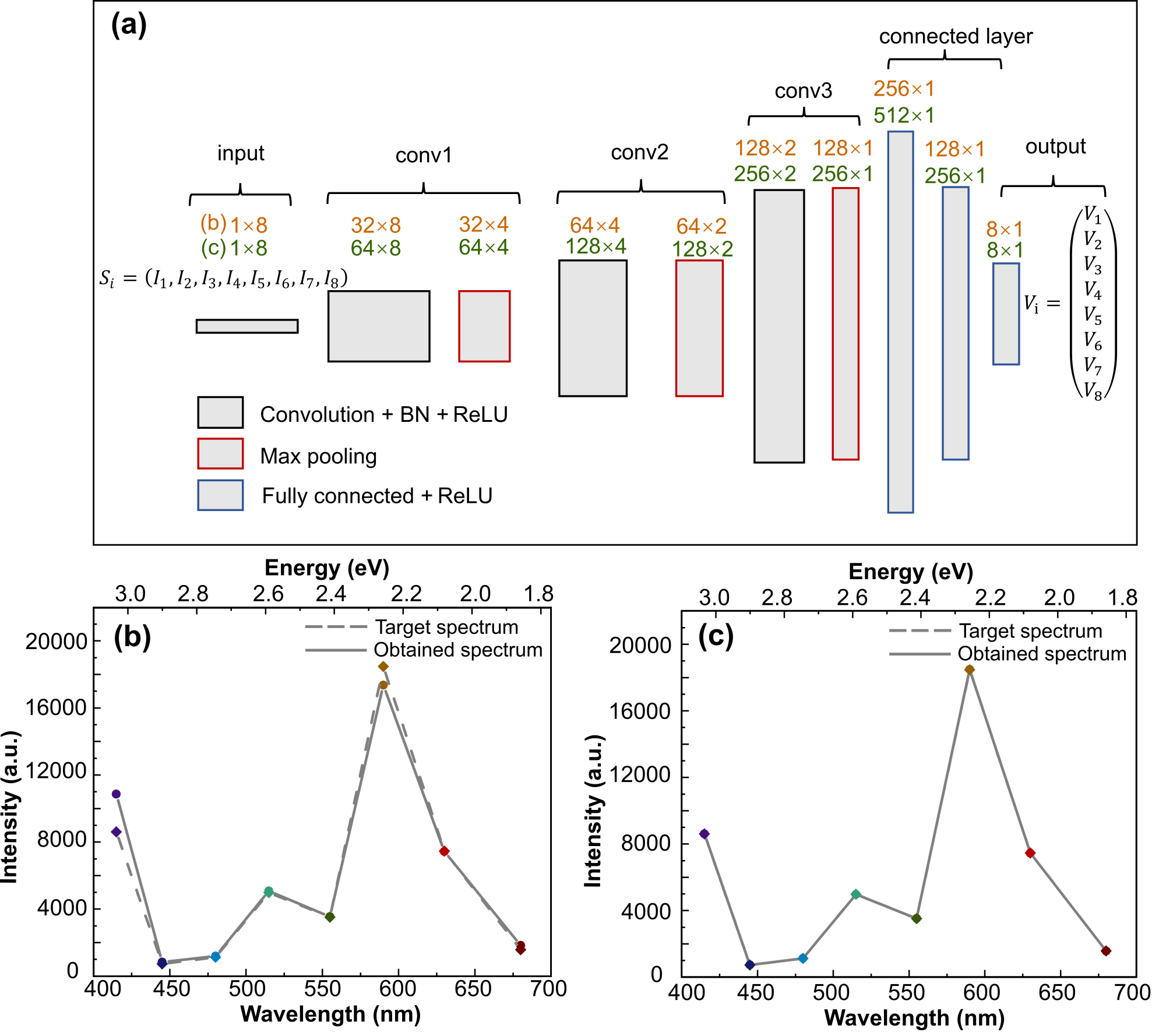}
  \caption{Deep-learning controller for closed-loop spectral optimization.  
  \textbf{(a)}~Convolutional neural network (CNN) architecture designed to regress optimal LED voltages from target spectra. The network comprises three convolutional layers with batch normalization and ReLU activation, interleaved with max pooling, followed by fully connected layers. Two filter configurations were evaluated: 32–64–128 and 64–128–256.  
  \textbf{(b)}~Inference result obtained with the 32–64–128 configuration, showing partial alignment with the target spectrum.  
  \textbf{(c)}~Inference result obtained with the deeper 64–128–256 configuration, demonstrating significantly improved agreement with the target spectrum and superior generalization capability.}
  \label{fig:6}
\end{figure*}

Experiments were conducted in a light‑tight enclosure; each algorithm followed five steps: sensor calibration, random voltage application, ML inference, spectrum comparison and iteration until convergence (max 400 steps).

%=====================================================================
\section{Results and Discussion}
\label{sec:results}

\subsection{Algorithm Performance}

The implementation of the three algorithms (traversal, Bayesian inference, and deep learning) provides complementary strategies for the design of autonomous closed-loop systems in spectral optimization.  Each method was assessed by its ability to identify the optimal set of LED voltages required to reproduce a predefined target spectrum, as quantified by the discrepancy between measured and reference outputs.  

Figure~\ref{fig:3} highlights the workflow and performance of the traversal algorithm.  
The schematic in Fig.~\ref{fig:3}(a) illustrates its hierarchical, tree-based exploration of the voltage parameter space, where incremental updates are applied iteratively across the LED channels to minimize spectral mismatch.  
This brute-force strategy leverages local comparisons and directional heuristics to gradually approach the optimum solution.  

The effect of voltage resolution on convergence is evident in Figs.~\ref{fig:3}(b)–\ref{fig:3}(c).  
With a coarse step size of 0.1\,V, optimization proceeds rapidly but yields only partial agreement with the target due to insufficient sampling of the voltage space.  
By contrast, a finer resolution of 0.05\,V improves spectral fidelity at the expense of longer iteration times.  

Although conceptually simple and straightforward to implement, the traversal algorithm is inherently sensitive to initialization choices and measurement noise.  
Robust performance therefore requires careful adjustment of hyperparameters, including step size and convergence thresholds.  
While not optimal for high-precision tasks involving nonlinear responses, traversal serves as a valuable baseline method for benchmarking algorithmic performance in low-resource or hardware-constrained environments.\\

Figure~\ref{fig:4} presents the working principle and iterative refinement process of the Bayesian optimization algorithm within the closed-loop autonomous platform.  
As shown in Fig.~\ref{fig:4}(a), the algorithm constructs a probabilistic surrogate model $\mu(x)$ from sparse initial data, capturing both the mean prediction and its associated uncertainty through Gaussian process regression.  
With successive iterations, illustrated in Fig.~\ref{fig:4}(b), the surrogate model is progressively refined by incorporating additional voltage–spectrum pairs, which reduces predictive uncertainty and improves accuracy.  
By iteration 400 [Fig.~\ref{fig:4}(c)], the predicted and measured spectra are nearly indistinguishable, indicating convergence of the surrogate model to the underlying system response.  

Bayesian optimization provides important advantages over deterministic methods because it explicitly accounts for uncertainty, which makes it particularly effective in the presence of noisy measurements or costly evaluations.  
The gradual narrowing of posterior variance enables efficient convergence with relatively few iterations, demonstrating the suitability of Bayesian optimization for practical applications in autonomous spectroscopy and other adaptive experimental systems.\\

Figure~\ref{fig:5} illustrates the data preparation and preprocessing pipeline used to train the deep learning controller.  
As shown in Fig.~\ref{fig:5}(a), each LED is driven independently under varying voltages, and the corresponding spectral intensities are recorded using an eight-channel photodetector.  
These single-LED responses form the foundation for constructing composite spectra.  
Synthetic data samples are then generated by randomly combining voltage vectors and summing the corresponding single-LED spectral outputs [Fig.~\ref{fig:5}(b)].  
This augmentation strategy provides a large and diverse training dataset while reducing experimental time.  
Both voltage vectors and spectral outputs are subsequently normalized [Fig.~\ref{fig:5}(c)] to ensure consistent scaling and to enhance training stability.  
The final dataset [Fig.~\ref{fig:5}(d)] contains approximately 100,000 voltage–spectrum pairs that serve as input for supervised learning.  

This scalable data generation framework enables efficient preparation of large and structured datasets that are essential for neural network training, particularly in situations where direct physical sampling is limited by time, energy, or hardware constraints. \\

Figure~\ref{fig:6} presents the predictive performance of the convolutional neural network (CNN) controller trained on the synthetic dataset.  
The network architecture [Fig.~\ref{fig:6}(a)] consists of three convolutional layers with batch normalization, ReLU activations, and max pooling, followed by fully connected layers that regress the optimal voltage vector from input spectral targets.  
Two filter configurations were evaluated: 32–64–128 and 64–128–256.  
The results in Fig.~\ref{fig:6}(b) and Fig.~\ref{fig:6}(c) show that the deeper configuration (64–128–256) achieves closer agreement between predicted and target spectra, confirming the effectiveness of increased model capacity.  

The CNN-based controller achieves high accuracy and rapid inference, which makes it a strong candidate for real-time, high-dimensional control tasks.  
Although the approach requires a substantial offline training phase, the trained model demonstrates robust generalization across test cases and provides both precision and scalability for closed-loop experimental systems.  

Overall, the deep learning algorithm is well suited for modeling the complex and nonlinear interactions between sensor readings and LED voltages.  
Despite its higher demand for training data and computational resources, it consistently outperforms traversal and Bayesian approaches in terms of accuracy and robustness.\\

\subsection{Comparison of Methods}

A comparative analysis underscores the distinct strengths and limitations of each algorithm.  
The traversal method is valued for its simplicity and low computational cost, but it converges slowly and is highly sensitive to initial conditions.  
The Bayesian approach performs well in noisy environments and provides the additional benefit of uncertainty quantification through posterior distributions, although it requires more complex computation and careful calibration of prior parameters.  
Deep learning is particularly effective in capturing nonlinear relationships within the data.  
Although the training phase is resource-intensive, once trained the model infers optimal voltages rapidly, making it highly suitable for real-time operation.  \\

A detailed comparison of traversal, Bayesian, and deep learning algorithms in terms of speed and accuracy is provided in Table~\ref{tab:algorithms}.

% ---- Table II: Runtime & accuracy comparison of algorithms ----
\begin{table*}[htbp]
\caption{\label{tab:algorithms} Comparison of traversal, Bayesian, and CNN algorithms in terms of speed and accuracy.}
\begin{threeparttable}
\begin{ruledtabular}
\begin{tabular}{p{1.5cm} p{4cm} p{3.1cm} p{2cm} p{1.5cm}}
\textbf{Algorithm} & \textbf{Description} & \multicolumn{2}{c}{\textbf{Time per spectrum (s)}} & \textbf{Error} \\
\cline{3-4}
 & & \textbf{Optimization (training for CNN)} & \textbf{Prediction} & \\
\hline
Traversal1 & Step size of 0.1V & \multicolumn{2}{c}{397.8} & 3812 \\
Traversal2 & Step size of 0.05V & \multicolumn{2}{c}{2861.5} & 257 \\
Bayes & 10 iterations & \multicolumn{2}{c}{27.1} & 3536 \\
Bayes & 400 iterations & \multicolumn{2}{c}{1119.4} & 359 \\
CNN1 & Conv. layers: 32, 64, 128 & 8696.6 & 0.0135 & 2539 \\
CNN2 & Conv. layers: 64, 128, 256 & 2760.9 & 0.0069 & 98 \\
\end{tabular}
\end{ruledtabular}
\end{threeparttable}
\end{table*}

\subsection{Educational Implications}
Embedding these algorithms within an affordable IoT framework provides substantial educational value.  
Students can directly experience how theoretical principles of algorithmic control are applied to experimental systems, thereby enhancing critical thinking and problem-solving skills.  
The demonstrations in Figs.~\ref{fig:3}–\ref{fig:6} illustrate the characteristic behavior and performance of each algorithm, making them effective instructional tools in applied physics and engineering curricula.  
By integrating such experiments into coursework, students are better prepared for interdisciplinary challenges and are encouraged to develop innovative solutions using AI- and IoT-based technologies.

\subsection{Limitations and Future Directions}
Although the present platform demonstrates robust closed-loop operation, several limitations merit consideration.  
The spectral resolution and dynamic range of the AS7341 sensor constrain measurement accuracy relative to high-end spectrometers.  
Hardware stability, including thermal drift of LEDs and susceptibility to ambient light, may also affect reproducibility.  
Future improvements could involve integration of higher-resolution detectors, active source calibration, and advanced hybrid control strategies that combine traversal, Bayesian inference, and deep learning to balance accuracy, speed, and computational efficiency.  
In addition, the broader development of safe and reliable self-driving laboratories requires careful consideration of reproducibility, robustness, and ethical deployment of autonomous experimentation~\cite{Leong2025}.  
Equally important are engineering principles such as process intensification, which can enhance the efficiency and scalability of SDLs across multiscale chemical and physical domains~\cite{DelgadoLicona2025}.

Such developments would extend the platform’s utility beyond instructional demonstrations toward broader applications in autonomous materials discovery and device prototyping.

%---------------------------------------------------------------------
\section{Conclusion}
\label{sec:conclusion}

In this study, we implemented and compared three algorithms—traversal, Bayesian and deep learning—within an affordable Internet‑of‑Things framework designed for self‑driving experiments. Using a setup featuring eight LEDs of different wavelengths and a corresponding light‑sensor array, our goal was to accurately infer the correct voltages for the LEDs by aligning sensor data (spectra) with target spectra.

The traversal algorithm demonstrated simplicity and low computational requirements, making it suitable for resource‑constrained applications. However, its performance was sensitive to initial conditions and noise, leading to potential inaccuracies. The Bayesian algorithm excelled in managing measurement noise and provided robust voltage estimations with quantified uncertainties by employing probabilistic reasoning. Meanwhile, the deep learning algorithm outperformed both in capturing complex, nonlinear relationships, achieving superior accuracy, albeit with higher demands for training data and computational resources.

Our results suggest that the choice of algorithm should align with specific application needs, such as the desired level of accuracy, available computational resources and system complexity. The traversal algorithm is well‑suited for rapid deployment with limited resources; the Bayesian algorithm is ideal where uncertainty quantification is crucial; and the deep learning approach is optimal for high‑accuracy, data‑intensive systems. Future research could explore hybrid approaches that integrate these algorithms to leverage their strengths and extend the system’s scalability and functionality.

\section*{Supplementary Material}
Supplementary material consists of two parts. 
First, a laboratory guide provides step-by-step instructions for assembling the IoT-based optical platform, programming the Arduino interface, and implementing traversal, Bayesian, and deep learning algorithms for voltage inference. This guide is intended to support both reproducibility of the experiments and their integration into physics education. 
Second, supporting information includes additional figures and analyses, such as detailed system components, 3D-printed fixtures, shielding conditions, LED stability tests, convergence behavior of the algorithms, dataset preparation procedures, and sensor validation results. 
Together, these materials complement the main manuscript and enable full reproduction of the experiments.

\begin{acknowledgments}
 This work received funding from the National Natural Science Foundation of China (NSFC) Grants (No. 62011530438, No. 61704129) are acknowledged. This work was partially supported by Fundamental Research Funds for the Central Universities (QTZX23026), the fund of the State Key Laboratory of Solidification Processing in Northwestern Polytechnical University (grant no. SKLSP201612), and the Open
Fund of State Key Laboratory of Infrared Physics (Grant No.
SITP-NLIST-ZD-2024-01).Y.X. acknowledgments the European Research Council through the ERC-2024-PoC StEnSo (Grant Agreement 101185235) and the ERC-2024-SyG SKIN2DTRONICS (grant agreement 101167218). We would also like to acknowledge the {\em Severo Ochoa Centres of Excellence\/} program through Grant CEX2024-001445-S. The authors also acknowledge Dr. Eduardo R. Hernandez (ICMM,CSIC) and Dr. Andres Castellanos-Gomez (ICMM,CSIC) for the careful reading of the manuscript. 

\end{acknowledgments}

\section*{Data and Code Availability}

All source code, Arduino firmware, and representative datasets supporting this work are openly available via Zenodo~\cite{Xie2025_Zenodo} (\href{https://doi.org/10.5281/zenodo.16944622}{DOI: 10.5281/zenodo.16944622}), which provides a permanent, citable archive of the exact version used in this study.

In addition, the actively maintained development repository is hosted on GitHub (\url{https://github.com/YongXie-ICMM/APL_Machine_Learning_2025}), where users can access the latest updates, contribute improvements, and track ongoing development.

The repository contains the complete workflow, including Python control scripts, dataset generation procedures, and trained models, enabling full reproducibility of the experiments and facilitating broader adoption of the platform in both educational and research contexts.

% --- Bibliography ---
\bibliographystyle{aipnum4-1} % numerical, sorted & compressed with above setcitestyle
\bibliography{aipsamp}

\end{document}